# Reducing the Computational Complexity of Multicasting in Large-Scale Antenna Systems

Meysam Sadeghi, *Student Member, IEEE*, Luca Sanguinetti, *Senior Member, IEEE*, Romain Couillet, *Senior Member, IEEE*, and Chau Yuen, *Senior Member, IEEE*

*Abstract*—In this paper, we study the physical layer multicasting to multiple co-channel groups in large-scale antenna systems. The users within each group are interested in a common message and different groups have distinct messages. We aim at designing the precoding vectors solving the so-called quality of service (QoS) and weighted max-min fairness (MMF) problems, assuming that the channel state information is available at the base station (BS). To solve both problems, the baseline approach exploits the semidefinite relaxation (SDR) technique. Considering a BS with $N$ antennas, the SDR complexity is more than $\mathcal{O}(N^6)$, which prevents its application in large-scale antenna systems. To overcome this issue, we present two new classes of algorithms that, not only have significantly lower computational complexity than existing solutions, but also largely outperform the SDR based methods. Moreover, we present a novel duality between transformed versions of the QoS and the weighted MMF problems. The duality explicitly determines the solution to the weighted MMF problem given the solution to the QoS problem, and vice versa. Numerical results are used to validate the effectiveness of the proposed solutions and to make comparisons with existing alternatives under different operating conditions.

## I. INTRODUCTION

The advent of data-hungry services and applications has significantly increased the amount of data traffic of wireless networks [1]. A considerable amount of this traffic belongs to the services that are of interest to one or several groups of subscribers such as news headlines, financial data, regular system updates, and video broadcasting [1], [2]. The traditional unicast technology is highly inefficient for these services as it ignores the nature of such a traffic demand [2]–[4]. To address this issue, the multicasting technology has been included in different releases of the third generation partnership project (3GPP) [2].

Physical layer multicasting is an efficient multicasting technique designed for wireless networks [3], [4]. It has been widely studied in the literature either for single or multiple groups of users [3]–[14]. In single-group multicasting, a transmitter exploits channel state information (CSI) to send out a common stream of data to one group of users, while in multigroup multicasting multiple independent streams of common data are sent to multiple distinct groups of users. In this context, two classes of problems have received particular attention, the so-called quality-of-service (QoS) problem and the weighted max-min-fairness (MMF)[1] problem. The former aims to minimize the total transmit power while satisfying target signal-to-interference-plus-noise ratios (SINRs) at the active user equipments (UEs). The latter seeks to maximize the minimum weighted SINR among all the UEs in the system, subject to a total transmit power constraint.

A seminal treatment of single-group multicasting for both QoS and MMF problems was first presented in [3]. Therein, it was proved that both QoS and MMF problems are NP-hard and then an approximate solution was presented employing semidefinite relaxation (SDR) technique [15]. This work is then extended to a multigroup single-cell scenario in [4]. It should be noted that, since in the multigroup case the SINR of every UE depends on the precoding vectors of all other groups, even finding a feasible solution for QoS and MMF problems might be a challenging task [4], [11]. Therefore, in [4] the SDR technique is followed by a randomization and a multigroup multicast power control phase. In [5], the MMF problem is studied under per-antenna power constraint for multigroup single-cell systems. The coordinated physical layer multicasting for single-group multicell scenario is investigated in [6]. Also, its application to coordinated physical layer multicasting for multigroup multicell scenario is studied in [14].

The aforementioned works (among many others) are based on the SDR technique, which is characterized by high computational complexity when the system dimensions grow large, especially for large antenna arrays. More precisely, consider a single-cell network wherein a BS with $N$ antennas serves $K$ UEs in $G$ multicasting groups. Then, solving the QoS problem via SDR requires $\mathcal{O}(\sqrt{GN})$ iterations of an interior point method with each iteration requiring $\mathcal{O}(G^3 N^6 + KGN^2)$ arithmetic operations [4]. The computational cost of finding an approximate solution for the MMF problem is even higher as its solution is achieved by iteratively solving different instances of the QoS problem. Therefore, the SDR-based solu-

M. Sadeghi (meysam@mymail.sutd.edu.sg) and C. Yuen (yuenchau@sutd.edu.sg) are with Singapore University of Technology and Design (SUTD), Singapore. L. Sanguinetti (luca.sanguinetti@unipi.it) is with the University of Pisa, Dipartimento di Ingegneria dell'Informazione, Italy and also with the Large Systems and Networks Group (LANEAS), CentraleSupélec, Université Paris-Saclay, France. R. Couillet (romain.couillet@centralesupelec.fr) is with the Signals and Statistics Group at CentraleSupélec, Université Paris-Saclay, France.

Part of the work is supported by A*Star SERC project number 142-02-00043. L. Sanguinetti and R. Couillet have been supported by the ERC Starting Grant 305123 MORE. L. Sanguinetti is also funded by the 5GIOTTO project from University of Pisa.

[1]For brevity, hereafter, we refer to the weighted MMF problem as MMF problem.

tions are not suitable for practical implementation when $N$, $G$, or $K$, grow large, as envisioned in large-scale antenna systems (commonly known as Massive MIMO systems) wherein $N$ can be of the order of hundreds [16]–[23].

In the context of Massive MIMO multicasting, two possible approaches have been recently proposed, namely, the asymptotic approach and the successive convex approximation (SCA) approach [11], [23]–[28]. The former exploits the asymptotic orthogonality of the channel vectors, when $N$ grows very large and $K$ is kept fixed, to simplify the SINR expression of each UE and facilitate the design of asymptotically optimal beamforming schemes [23]–[25]. In particular, [25] investigates the MMF problem for the multigroup single-cell multicasting whereas the single-group multicell case is studied in [24]. The extension to a multigroup multicell network is considered in [23]. The main problem with the asymptotic approach is that an extremely large number of antennas is required to reach the asymptotic orthogonality condition. As a consequence, the performance of the asymptotically optimal precoders is poor when the system does not have an extremely large number of antennas (in the order of thousands) [23].

The SCA approach aims at iteratively solving the non-convex QoS and MMF problems by means of SCA of the original problems around a feasible point [29], [30]. More specifically, the algorithm starts from an initial feasible point, the non-convex constraints are approximated by convex functions around this point, and the resulting convex problem is solved before proceeding to the next iteration. This procedure is repeated until convergence to a stationary point. In [26], the SCA has been applied to reduce the computational complexity of beamforming design in single-group multicasting for large-scale antenna arrays. However, the SCA method is not suitable for multigroup multicasting communications as it requires an initial feasible point, which is hard to compute in these scenarios [11]. To handle this issue, a feasible point pursuit SCA (FPP-SCA) algorithm is proposed in [27] and applied to multigroup multicasting in [11]. Therein, in order to guarantee the feasibility of the problem, slack variables are added to relax the constraints, and a penalty is used to ensure that slacks are sparingly used. The solution of the resulting optimization problem is then used for another round of approximation and the procedure is repeated until convergence. However, the method itself has two drawbacks. First, although the solution of the approximated problem is always feasible, it might not be a feasible solution of the original multicasting problem and it is sensitive to the initial point of the algorithm as it is detailed in [27]. Second, it is still computationally demanding when the number of antennas grows.

In this paper, we address all the aforementioned drawbacks for a multigroup single-cell system by introducing a two-layer precoding scheme, which is tailored for large-scale antenna systems. Our main contributions are summarized as follows:

1) We present two algorithms for the QoS and MMF problems, that outperform most of the aforementioned solutions while guaranteeing a low computational complexity.

2) We reveal new duality results that allow to solve both QoS and MMF problems simultaneously. This is in sharp contrast with the existing algorithms for which the MMF problem is solved by iteratively solving different instances of the QoS problem.

3) We introduce a heuristic algorithm that significantly improves the computational complexity while only slightly reducing the performance of both QoS and MMF solutions.

The remainder of the paper is organized as follows. Section II introduces the system model for a multigroup single-cell large-scale antenna array system and formulates the corresponding QoS and MMF problems. Section III introduces the proposed two-layer precoder, it provides a duality result between transformed versions of the QoS and MMF problems, and then it proposes two algorithms for solving both. Section IV introduces a heuristic solution to further reduce the computational complexity. Section V presents the numerical results whereas conclusions are drawn in Section VI.

*Notations:* Scalars are denoted by lower case letters whereas boldface lower (upper) case letters are used for vectors (matrices). We denote by $\mathbf{0}$ a matrix of appropriate size where all its elements are zero. The transpose, conjugate transpose, real part, absolute value, and second norm operator are denoted by $(\cdot)^T$, $(\cdot)^H$, $\mathrm{Re}(\cdot)$, $|\cdot|$, and $\|\cdot\|$. The set of all positive real numbers is denoted by $\mathbb{R}^+$. A circular symmetric complex Gaussian random vector $\mathbf{x}$ is denoted by $\mathbf{x} \sim \mathcal{CN}(\mathbf{0}, \mathbf{C})$, where $\mathbf{0}$ and $\mathbf{C}$ are its mean and covariance matrix, respectively. The inverse of an invertible function $f(.)$ is shown by $f^{-1}(.)$.

## II. System Model and Problem Formulation

Consider a single-cell large-scale antenna array system in which a BS equipped with $N$ antennas serves $G$ multicasting groups. Denote by $\mathcal{G} = \{1, \ldots, G\}$ the set of indices of all groups and call $\mathcal{K}_j$ the set of UE indices associated with group $j$, with cardinality $K_j = |\mathcal{K}_j|$ and such that $\mathcal{K}_j \cap \mathcal{K}_i = \emptyset$, $j \neq i$, i.e., each UE is associated with a single group. Within this setting, we assume that $N > K - \min_{j \in \mathcal{G}} K_j$, where $K = \sum_{j=1}^{G} K_j$ is the total number of UEs in the network. Since we consider large antenna systems, this technical assumption is naturally in place. A double index notation is used to refer to each UE as e.g., "user $k$ in group $j$". Under this convention, let $\mathbf{g}_{jk} \in \mathbb{C}^N$ be the channel between UE $k$ in group $j$ and the BS and assume that $\mathbf{g}_{jk} = \sqrt{\beta_{jk}}\mathbf{h}_{jk}$, where $\mathbf{h}_{jk} \sim \mathcal{CN}(\mathbf{0}_N, \mathbf{I}_N)$ is the small-scale fading channel and $\beta_{jk}$ accounts for the large-scale channel attenuation (or path loss). We assume that the BS has perfect knowledge of the channel vectors $\{\mathbf{g}_{jk}\}$.

Denoting by $\mathbf{w}_j \in \mathbb{C}^N$ the precoding vector associated with group $j$, the signal $y_{jk}$ received at UE $k$ can be written as:

$$y_{jk} = \mathbf{g}_{jk}^H \mathbf{w}_j s_j + \sum_{i=1, i \neq j}^{G} \mathbf{g}_{jk}^H \mathbf{w}_i s_i + n_{jk} \qquad (1)$$

where $s_i \sim \mathcal{CN}(0, 1)$ is the signal intended to group $i$, assumed independent across $i$, and $n_{jk} \sim \mathcal{CN}(0, \sigma_{jk}^2)$ accounts for the additive Gaussian noise. The SINR $\gamma_{jk}$ of UE $k$ in group $j$ can be written as

$$\gamma_{jk} = \frac{|\mathbf{g}_{jk}^H \mathbf{w}_j|^2}{\sum_{i=1,i\neq j}^G |\mathbf{g}_{jk}^H \mathbf{w}_i|^2 + \sigma_{jk}^2} \qquad (2)$$

and the total average transmit power is $\sum_{j=1}^G \|\mathbf{w}_j\|^2$. Under the above assumptions, an instance of the QoS problem can be formulated as follows [4]:

$$\mathcal{Q}(\boldsymbol{\eta}): \quad \min_{\{\mathbf{w}_j\}} \quad \sum_{j=1}^G \|\mathbf{w}_j\|^2 \qquad (3)$$

$$\text{s.t.} \quad \gamma_{jk} \geq \eta_{jk} \quad \forall j,k \qquad (4)$$

where $\eta_{jk}$ is the prescribed SINR of UE $k$ in group $j$ and $\boldsymbol{\eta} \in \mathbb{C}^K$ is the vector collecting all the $\{\eta_{jk}\}$. Accordingly, an instance of the MMF problem is [4]:

$$\mathcal{F}(\boldsymbol{\eta}, P): \quad \max_{\{\mathbf{w}_j\}} \min_j \min_k \frac{1}{\eta_{jk}} \gamma_{jk} \qquad (5)$$

$$\text{s.t.} \quad \sum_{j=1}^G \|\mathbf{w}_j\|^2 \leq P \qquad (6)$$

where $P$ accounts for the power constraint at the BS, and $\frac{1}{\eta_{jk}}$ represents the weight of $\gamma_{jk}$. As mentioned before, $\mathcal{Q}(\boldsymbol{\eta})$ and $\mathcal{F}(\boldsymbol{\eta}, P)$ are NP-hard and the existing algorithms for computing their approximate solutions have either high computational complexity [3], [5], [11], or poor performance [23], [24], [28]. A two-layered architecture is proposed next to overcome these drawbacks.

## III. THE PROPOSED TWO-LAYER PRECODING SCHEME

In this section, we propose a simple and computationally efficient method to compute approximate solutions to the QoS and MMF problems. The method is based on a two-layer precoding scheme: (i) the outer layer restricts the space of valid precoders to those cancelling the inter-group interference, thereby approximating the QoS and MMF problems by simpler (still non-convex) problems, denoted by QoS$_{\text{dec}}$ and MMF$_{\text{dec}}$, for which trivial feasible points can be found; (ii) starting from these feasible points, the inner layer is designed to reach a suboptimal solution to the QoS$_{\text{dec}}$ and MMF$_{\text{dec}}$ problems, which are also feasible solutions of the original QoS and MMF problems. Section III-A presents the outer layer. Section III-B reveals an explicit duality between the QoS$_{\text{dec}}$ and MMF$_{\text{dec}}$ problems. Section III-C presents the inner layer and the algorithms developed. Section III-D evaluates the complexity of the proposed algorithms.

### A. Outer Layer – Removing Multigroup Interference

Denote by $\mathbf{G}_i \in \mathbb{C}^{N \times K_i}$ the matrix collecting the channel vectors of all the $K_i$ UEs in group $i$. The complete elimination of the multigroup interference $\sum_{i=1,i\neq j}^G \mathbf{g}_{jk}^H \mathbf{w}_i s_i$ in (1) is achieved by using the block-diagonalization zero-forcing (BDZF) technique [31], [32]. Consider a two-layer precoding vector for group $j$ as follows

$$\mathbf{w}_j = \mathbf{F}_j \mathbf{c}_j \qquad \forall j \in \mathcal{G} \qquad (7)$$

where $\mathbf{c}_j \in \mathbb{C}^{N-\tau_j}$ with $\tau_j = K - K_j$ is the inner layer, the design of which is discussed later, and $\mathbf{F}_j \in \mathbb{C}^{N \times (N-\tau_j)}$ is the outer layer. We design $\mathbf{F}_j$ as an isometric matrix whose columns form a basis for the null space of $\mathbf{G}_{-j} = [\mathbf{G}_1, \ldots, \mathbf{G}_{j-1}, \mathbf{G}_{j+1}, \ldots, \mathbf{G}_G] \in \mathbb{C}^{N \times \tau_j}$, i.e., $\mathbf{G}_{-j}^H \mathbf{F}_j = \mathbf{0}_{\tau_j \times (N-\tau_j)}$. As proposed in [31], [32], $\mathbf{F}_j$ can be obtained through the singular value decomposition (SVD) of $\mathbf{G}_{-j}$. This requires $16(G-1)KN^2 + 24N\sum_{j=1}^G (K-K_j)^2$ floating point operations (flops)[2]. The same goal can be obtained with lower complexity (linear in the number of BS antennas $N$) using the QR-based decomposition approach as shown in [34]. This produces

$$\mathbf{G}_{-j} = \mathbf{Q}_j \mathbf{R}_j = \begin{bmatrix} \mathbf{Q}_j' & \mathbf{Q}_j'' \end{bmatrix} \begin{bmatrix} \mathbf{R}_j' \\ \mathbf{0} \end{bmatrix} = \mathbf{Q}_j' \mathbf{R}_j' \qquad (8)$$

where $\mathbf{Q}_j'' \in \mathbb{C}^{N \times (N-\tau_j)}$ gives the null space of $\mathbf{G}_{-j}$ such that $\mathbf{G}_{-j}^H \mathbf{Q}_j'' = \mathbf{0}_{\tau_j \times (N-\tau_j)}$. Therefore we can use $\mathbf{Q}_j''$ as the outer layer of $\mathbf{w}_j$, i.e., $\mathbf{F}_j = \mathbf{Q}_j''$. Since the QR decomposition of an $m$ by $n$ matrix can be computed with $8mn^2 - 8/3n^3$ flops [33], the total number of flops required to perform the BDZF technique reduces to $8N\sum_{j=1}^G (K-K_j)^2 - 8/3\sum_{j=1}^G (K-K_j)^3$, which increases linearly with $N$. Plugging $\mathbf{F}_j = \mathbf{Q}_j''$ into (2) yields

$$\gamma_{jk} = |\overline{\mathbf{g}}_{jk}^H \mathbf{c}_j|^2 \qquad (9)$$

where $\overline{\mathbf{g}}_{jk} = \frac{1}{\sigma_{jk}} (\mathbf{Q}_j'')^H \mathbf{g}_{jk} \in \mathbb{C}^{N-\tau_j}$ denotes the equivalent channel vector of UE $k$ in group $j$. As $\forall j$ $(\mathbf{F}_j)^H \mathbf{F}_j = \mathbf{I}_{N-\tau_j}$, the proposed outer layer does not change the transmit power, i.e., $\sum_{j=1}^G \|\mathbf{w}_j\|^2 = \sum_{j=1}^G \|\mathbf{c}_j\|^2$. Therefore, using the BDZF technique the QoS problem reduces to QoS$_{\text{dec}}$. Note that we define the QoS$_{\text{dec}}$ as the transformed version of the QoS problem into $G$ single-group multicasting QoS problems, $\{\overline{\mathcal{Q}}_j(\boldsymbol{\eta}_j)\}_{j=1}^G$, where the $j$th problem is given by

$$\overline{\mathcal{Q}}_j(\boldsymbol{\eta}_j): \quad \min_{\{\mathbf{c}_j\}} \quad \|\mathbf{c}_j\|^2 \qquad (10)$$

$$\text{s.t.} \quad |\overline{\mathbf{g}}_{jk}^H \mathbf{c}_j|^2 \geq \eta_{jk} \quad \forall k \qquad (11)$$

where $\boldsymbol{\eta}_j \in \mathbb{C}^{K_j}$ is the vector collecting all the quantities $\{\eta_{jk}\}$ in group $j$. To grasp the relation between the QoS$_{\text{dec}}$ problem and the prescribed SINRs, i.e., $\{\eta_{jk}\}_{j=1}^G$, we denote an instance of the QoS$_{\text{dec}}$ problem by $\overline{\mathcal{Q}}(\boldsymbol{\eta}) = \{\overline{\mathcal{Q}}_j(\boldsymbol{\eta}_j)\}_{j=1}^G$. Accordingly, using the BDZF technique, the MMF problem reduces to MMF$_{\text{dec}}$, where we show an instance of it by $\overline{\mathcal{F}}(\boldsymbol{\eta}, P)$ and it is given as follows

$$\overline{\mathcal{F}}(\boldsymbol{\eta}, P): \quad \max_{\{\mathbf{c}_j\}} \min_j \min_k \frac{1}{\eta_{jk}} |\overline{\mathbf{g}}_{jk}^H \mathbf{c}_j|^2 \qquad (12)$$

$$\text{s.t.} \quad \sum_{j=1}^G \|\mathbf{c}_j\|^2 \leq P. \qquad (13)$$

As mentioned in the introduction, finding a feasible point for $\mathcal{Q}(\boldsymbol{\eta})$ is hard [4]. The same holds for $\mathcal{F}(\boldsymbol{\eta}, P)$ since the common approach to solve $\mathcal{F}(\boldsymbol{\eta}, P)$ relies on iteratively solving $\mathcal{Q}(\boldsymbol{\eta})$. On the contrary, finding a feasible point for $\overline{\mathcal{Q}}(\boldsymbol{\eta}) = \{\overline{\mathcal{Q}}_j(\boldsymbol{\eta}_j)\}$ and, thus, for $\overline{\mathcal{F}}(\boldsymbol{\eta}, P)$ is a straightforward task, thanks to the single-group nature of each subproblem $\overline{\mathcal{Q}}_j(\boldsymbol{\eta}_j)$. More specifically, for any given group $j$ an initial feasible point can be computed by first choosing an arbitrary beamforming vector $\mathbf{c}_j$ and then rescaling it so as to meet the most violated SINR constraint with equality. Despite being simpler to solve than $\mathcal{Q}(\boldsymbol{\eta})$ and $\mathcal{F}(\boldsymbol{\eta}, P)$, both $\overline{\mathcal{Q}}(\boldsymbol{\eta})$ and $\overline{\mathcal{F}}(\boldsymbol{\eta}, P)$ are still NP-hard – as it easily follows observing that for $G = 1$ they reduce to the single-group problems studied in [3]. The main difficulty in solving $\overline{\mathcal{Q}}(\boldsymbol{\eta})$ and $\overline{\mathcal{F}}(\boldsymbol{\eta}, P)$ lies in the non-convexity of the SINR constraints. In Section III-C, the SCA technique is used to develop a possible solution capable of overcoming this issue.

---

[2] The SVD calculation of an $m$ by $n$ matrix requires $16m^2 n + 24mn^2$ flops [33].

Before delving into this, next we further detail the characteristics of $\overline{\mathcal{Q}}(\boldsymbol{\eta})$ and $\overline{\mathcal{F}}(\boldsymbol{\eta}, P)$ and establish a duality and direct relations between the two problems: the solution of $\overline{\mathcal{F}}(\boldsymbol{\eta}, P)$ can be obtained from that of $\overline{\mathcal{Q}}(\boldsymbol{\eta})$ (and vice versa). These results will be used in Section III-C and Section IV to compute an approximate solution to $\overline{\mathcal{F}}(\boldsymbol{\eta}, P)$ by means of $\overline{\mathcal{Q}}(\boldsymbol{\eta})$ without the need of iteratively solving instances of $\overline{\mathcal{Q}}(\boldsymbol{\eta})$ as for existing alternatives [4]–[6].

### B. On the duality between $\overline{\mathcal{Q}}(\boldsymbol{\eta})$ and $\overline{\mathcal{F}}(\boldsymbol{\eta}, P)$

Let $\{\mathbf{c}_j^\star(\boldsymbol{\eta})\}$ and $P^\star(\boldsymbol{\eta})$ denote the set of optimal precoding vectors and the optimal objective value of $\overline{\mathcal{Q}}(\boldsymbol{\eta})$, respectively. Similarly, let $\{\mathbf{c}_j^\circ(\boldsymbol{\eta}, P)\}$ and $t^\circ(\boldsymbol{\eta}, P)$ denote the set of optimal precoding vectors and the optimal objective value of $\overline{\mathcal{F}}(\boldsymbol{\eta}, P)$. We then start providing the following result:

**Lemma 1.** *For $\overline{\mathcal{Q}}(\boldsymbol{\eta})$ and $\overline{\mathcal{F}}(\boldsymbol{\eta}, P)$ we have:*

$$\mathbf{c}_j^\star(\alpha\boldsymbol{\eta}) = \mathbf{c}_j^\circ\left(\boldsymbol{\eta}, P^\star(\alpha\boldsymbol{\eta})\right) \quad \forall j \in \mathcal{G} \tag{14}$$

*with $\alpha = t^\circ\left(\boldsymbol{\eta}, P^\star(\alpha\boldsymbol{\eta})\right)$. Also, we have that:*

$$\mathbf{c}_j^\circ(\boldsymbol{\eta}, P) = \mathbf{c}_j^\star\left(t^\circ(\boldsymbol{\eta}, P)\boldsymbol{\eta}\right) \quad \forall j \in \mathcal{G} \tag{15}$$

*with $P = P^\star\left(t^\circ(\boldsymbol{\eta}, P)\boldsymbol{\eta}\right)$.*

*Proof.* The proof proceeds by contradiction. First, notice that, by definition, $\{\mathbf{c}_j^\star(\alpha\boldsymbol{\eta})\}$ is a feasible solution of $\overline{\mathcal{F}}(\boldsymbol{\eta}, P^\star(\alpha\boldsymbol{\eta}))$ with an objective value equal to $\alpha$. Now, let us assume there exists a set of precoding vectors $\{\mathbf{c}_j^\circ(\boldsymbol{\eta}, P^\star(\alpha\boldsymbol{\eta}))\}$ for which $t^\circ(\boldsymbol{\eta}, P^\star(\alpha\boldsymbol{\eta})) > \alpha$. Clearly, $\{\mathbf{c}_j^\circ(\boldsymbol{\eta}, P^\star(\alpha\boldsymbol{\eta}))\}$ is also a feasible solution of $\overline{\mathcal{Q}}(\alpha\boldsymbol{\eta})$ for which all the SINR constraints are over satisfied. Hence, there exists a constant $\nu < 1$ such that $\{\nu\mathbf{c}_j^\circ(\boldsymbol{\eta}, P^\star(\alpha\boldsymbol{\eta}))\}$ meets all the SINR constraints of $\overline{\mathcal{Q}}(\alpha\boldsymbol{\eta})$ with equality while providing a smaller objective value than $P^\star(\alpha\boldsymbol{\eta})$. This, however, contradicts our assumption and proves that (14) is valid with $\alpha = t^\circ(\boldsymbol{\eta}, P^\star(\alpha\boldsymbol{\eta}))$. A similar line of reasoning can be used to prove (15). By definition the set of precoding vectors $\{\mathbf{c}_j^\circ(\boldsymbol{\eta}, P)\}$ is a feasible solution of $\overline{\mathcal{Q}}(t^\circ(\boldsymbol{\eta}, P)\boldsymbol{\eta})$ with an objective value equal to $P$. Let us assume there exists $\{\mathbf{c}_j^\star(t^\circ(\boldsymbol{\eta}, P)\boldsymbol{\eta})\}$ with $P^\star(t^\circ(\boldsymbol{\eta}, P)\boldsymbol{\eta}) < P$. Then, one can use the remaining power, $P - P^\star(t^\circ(\boldsymbol{\eta}, P)\boldsymbol{\eta})$ to rescale $\{\mathbf{c}_j^\star(t^\circ(\boldsymbol{\eta}, P)\boldsymbol{\eta})\}$ and improve $\overline{\mathcal{F}}(\boldsymbol{\eta}, P)$. This is in contradiction with our assumption and completes the proof. □

Also, the following lemma can be simply proved from the definition of $\overline{\mathcal{Q}}(\boldsymbol{\eta})$:

**Lemma 2.** *For $\overline{\mathcal{Q}}(\alpha\boldsymbol{\eta})$ and $\forall \alpha \in \mathbb{R}^+$ we have*

$$P^\star(\alpha\boldsymbol{\eta}) = \alpha P^\star(\boldsymbol{\eta}) \tag{16}$$

*and $\forall j \in \mathcal{G}$*

$$\mathbf{c}_j^\star(\alpha\boldsymbol{\eta}) = \sqrt{\alpha}\mathbf{c}_j^\star(\boldsymbol{\eta}). \tag{17}$$

We are now ready to state the following explicit duality between $\overline{\mathcal{Q}}(\boldsymbol{\eta})$ and $\overline{\mathcal{F}}(\boldsymbol{\eta}, P)$:

**Theorem 1.** *Given the set of optimal precoding vectors and the optimal objective value of $\overline{\mathcal{Q}}(\boldsymbol{\eta})$, i.e., $\{\mathbf{c}_j^\star(\boldsymbol{\eta})\}$ and $P^\star(\boldsymbol{\eta})$, the set of optimal precoding vectors and the optimal objective value of $\overline{\mathcal{F}}(\boldsymbol{\eta}, P)$, i.e. $\{\mathbf{c}_j^\circ(\boldsymbol{\eta}, P)\}$ and $t^\circ(\boldsymbol{\eta}, P)$, are determined as*

$$\mathbf{c}_j^\circ(\boldsymbol{\eta}, P) = \sqrt{\frac{P}{P^\star(\boldsymbol{\eta})}}\mathbf{c}_j^\star(\boldsymbol{\eta}) \quad \forall j \in \mathcal{G} \tag{18}$$

$$t^\circ(\boldsymbol{\eta}, P) = \frac{P}{P^\star(\boldsymbol{\eta})} \tag{19}$$

*and vice versa as*

$$\mathbf{c}_j^\star(\boldsymbol{\eta}) = \frac{1}{\sqrt{t^\circ(\boldsymbol{\eta}, P)}}\mathbf{c}_j^\circ(\boldsymbol{\eta}, P) \quad \forall j \in \mathcal{G} \tag{20}$$

$$P^\star(\boldsymbol{\eta}) = \frac{P}{t^\circ(\boldsymbol{\eta}, P)}. \tag{21}$$

*Proof.* Starting with (18) we have that

$$\sqrt{\frac{P}{P^\star(\boldsymbol{\eta})}}\mathbf{c}_j^\star(\boldsymbol{\eta}) \overset{(a)}{=} \mathbf{c}_j^\star\left(\frac{P}{P^\star(\boldsymbol{\eta})}\boldsymbol{\eta}\right) \overset{(b)}{=} \mathbf{c}_j^\circ\left(\boldsymbol{\eta}, P^\star\left(\frac{P}{P^\star(\boldsymbol{\eta})}\boldsymbol{\eta}\right)\right) \tag{22}$$

$$\overset{(c)}{=} \mathbf{c}_j^\circ\left(\boldsymbol{\eta}, \frac{P}{P^\star(\boldsymbol{\eta})}P^\star(\boldsymbol{\eta})\right) = \mathbf{c}_j^\circ(\boldsymbol{\eta}, P) \tag{23}$$

where $(a)$ follows from (17), $(b)$ holds because of (14) and $(c)$ is obtained using (16). The equality (19) follows from

$$P \overset{(a)}{=} P^\star\left(t^\circ(\boldsymbol{\eta}, P)\boldsymbol{\eta}\right) \overset{(b)}{=} t^\circ(\boldsymbol{\eta}, P)P^\star(\boldsymbol{\eta}) \tag{24}$$

where $(a)$ exploits $P = P^\star\left(t^\circ(\boldsymbol{\eta}, P)\boldsymbol{\eta}\right)$ (see Lemma 1), and $(b)$ is due to (16). The equality in (20) follows from replacing (19) in (18). □

Theorem 1 reveals the relation between the optimal precoding vectors and the optimal objective values of $\overline{\mathcal{Q}}(\boldsymbol{\eta})$ and $\overline{\mathcal{F}}(\boldsymbol{\eta}, P)$. However, as they are NP-hard, any arbitrary algorithm with polynomial complexity can provide an approximate set of precoding vectors, rather than the optimal one. Hence, it is interesting to establish a relation between the precoding vectors and the objective values of $\overline{\mathcal{Q}}(\boldsymbol{\eta})$ and $\overline{\mathcal{F}}(\boldsymbol{\eta}, P)$ while they are achieved from any arbitrary sub-optimal algorithm. This relation is given in Propositions 1 and 2.

**Proposition 1.** *Assume $\{\mathbf{c}_{j,\mathrm{app}}^\star(\boldsymbol{\eta})\}$ is a set of precoding vectors of $\overline{\mathcal{Q}}(\boldsymbol{\eta})$ and $P_{\mathrm{app}}^\star(\boldsymbol{\eta})$ is its associated objective value achieved by any arbitrary algorithm. Then, the set of precoding vectors $\{\sqrt{\frac{P}{P_{\mathrm{app}}^\star(\boldsymbol{\eta})}}\mathbf{c}_{j,\mathrm{app}}^\star(\boldsymbol{\eta})\}$ (or $\{\sqrt{\frac{P}{P_{\mathrm{app}}^\star(\boldsymbol{\eta})}}\mathbf{F}_j\mathbf{c}_{j,\mathrm{app}}^\star(\boldsymbol{\eta})\}$) is a feasible answer for $\overline{\mathcal{F}}(\boldsymbol{\eta}, P)$ (or $\mathcal{F}(\boldsymbol{\eta}, P)$), and provides an objective value $t_{app}^\circ(\boldsymbol{\eta}, P)$ such that $t_{app}^\circ(\boldsymbol{\eta}, P) \in [\frac{P}{P_{\mathrm{app}}^\star(\boldsymbol{\eta})}, \frac{P}{P^\star(\boldsymbol{\eta})}]$.*

*Proof.* Please refer to the Appendix A. $\qquad \square$

**Proposition 2.** *Assume* $\{\mathbf{c}_{j,\mathrm{app}}^{\circ}(\boldsymbol{\eta}, P)\}$ *is a set of precoding vectors of* $\overline{\mathcal{F}}(\boldsymbol{\eta}, P)$ *and* $t_{\mathrm{app}}^{\circ}(\boldsymbol{\eta}, P)$ *is its associated objective value achieved by any arbitrary algorithm. Then, the set of precoding vectors* $\{\frac{1}{\sqrt{t_{\mathrm{app}}^{\circ}(\boldsymbol{\eta}, P)}} \mathbf{c}_{j,\mathrm{app}}^{\circ}(\boldsymbol{\eta}, P)\}$ *(or* $\{\frac{1}{\sqrt{t_{\mathrm{app}}^{\circ}(\boldsymbol{\eta}, P)}} \mathbf{F}_j \mathbf{c}_{j,\mathrm{app}}^{\circ}(\boldsymbol{\eta}, P)\}$ *), is a feasible answer for* $\overline{\mathcal{Q}}(\boldsymbol{\eta})$ *(or* $\mathcal{Q}(\boldsymbol{\eta})$ *), and provides an objective value* $P_{\mathrm{app}}^{\star}(\boldsymbol{\eta})$ *such that* $P_{\mathrm{app}}^{\star}(\boldsymbol{\eta}) \in [\frac{P}{t^{\circ}(\boldsymbol{\eta}, P)}, \frac{P}{t_{\mathrm{app}}^{\circ}(\boldsymbol{\eta}, P)}]$.

*Proof.* Please refer to the Appendix B. $\qquad \square$

Note that the relation between the QoS and MMF problems was first discovered in [4], but it was not given in an explicit form. Therefore, the existing works in the literature, as [4]–[6], [11], solve the MMF problem by iteratively solving specific instances of the QoS problem. By virtue of the large number of antennas available in large-scale antenna systems and the BDZF technique, Theorem 1, Proposition 1, and Proposition 2, state that $\overline{\mathcal{F}}(\boldsymbol{\eta}, P)$ and $\overline{\mathcal{Q}}(\boldsymbol{\eta})$ (also $\mathcal{F}(\boldsymbol{\eta}, P)$ and $\mathcal{Q}(\boldsymbol{\eta})$) can be solved simultaneously. It is also interesting to observe that the upper bound of the objective value of $\overline{\mathcal{F}}(\boldsymbol{\eta}, P)$ achieved via Proposition 1 is equal to (19). Also, the lower bound of the objective value of $\overline{\mathcal{Q}}(\boldsymbol{\eta})$ achieved via Proposition 2 is equal to (21).

### C. Inner Layer – Successive Convex Approximation

In the sequel, the SCA technique is applied to solve $\overline{\mathcal{Q}}(\boldsymbol{\eta})$ and $\overline{\mathcal{F}}(\boldsymbol{\eta}, P)$. We begin with $\overline{\mathcal{Q}}(\boldsymbol{\eta})$, and rewrite $|\bar{\mathbf{g}}_{jk}^H \mathbf{c}_j|^2$ as

$$|\bar{\mathbf{g}}_{jk}^H \mathbf{c}_j|^2 = \mathbf{c}_j^H \mathbf{X}_{jk} \mathbf{c}_j \qquad (25)$$

where $\mathbf{X}_{jk} = \bar{\mathbf{g}}_{jk} \bar{\mathbf{g}}_{jk}^H$ is a rank-one positive semi-definite matrix. Thus, for any arbitrary vector $\mathbf{z}_j \in \mathbb{C}^{N-\tau_j}$ we have that $(\mathbf{c}_j - \mathbf{z}_j)^H \mathbf{X}_{jk} (\mathbf{c}_j - \mathbf{z}_j) \geq 0$ from which it follows

$$\mathbf{c}_j^H \mathbf{X}_{jk} \mathbf{c}_j \geq 2\Re(\mathbf{z}_j^H \mathbf{X}_{jk} \mathbf{c}_j) - \mathbf{z}_j^H \mathbf{X}_{jk} \mathbf{z}_j. \qquad (26)$$

Now, for any $\mathbf{z}_j$ the non-convex SINR constraint $\mathbf{c}_j^H \mathbf{X}_{jk} \mathbf{c}_j \geq \eta_{jk}$ can be replaced with a tighter convex constraint given by

$$2\Re(\mathbf{z}_j^H \mathbf{X}_{jk} \mathbf{c}_j) - \mathbf{z}_j^H \mathbf{X}_{jk} \mathbf{z}_j \geq \eta_{jk}. \qquad (27)$$

By replacing (11) with (27), we obtain

$$\widetilde{\mathcal{Q}}_j(\boldsymbol{\eta}_j, \mathbf{z}_j) : \min_{\{\mathbf{c}_j\}} \quad \|\mathbf{c}_j\|^2 \qquad (28)$$

$$\text{s.t.} \quad 2\Re(\mathbf{z}_j^H \mathbf{X}_{jk} \mathbf{c}_j) - \mathbf{z}_j^H \mathbf{X}_{jk} \mathbf{z}_j \geq \eta_{jk} \ \ \forall k \qquad (29)$$

which represents a convex approximation of $\overline{\mathcal{Q}}_j(\boldsymbol{\eta}_j)$ for a specific instance of $\mathbf{z}_j$. Now, we can introduce Algorithm 1 and its following proposition for the QoS problem.

---

**Algorithm 1** The QoS BDZF-SCA Algorithm

1: Compute $\mathbf{F}_j$ $\forall j \in \mathcal{G}$.
2: **for** $j = 1$ to $G$ **do**
3:     Select an arbitrary $\mathbf{z}_j^{(1)}$ and rescale it such that $\forall k$ $\mathbf{z}_j^{(1)H} \mathbf{X}_{jk} \mathbf{z}_j^{(1)} \geq \eta_{jk}$.
4:     **repeat**
5:         Solve:

$$\widetilde{\mathcal{Q}}_j(\boldsymbol{\eta}_j, \mathbf{z}_j^{(i)}) : \min_{\mathbf{c}_j^{(i)}} \quad \|\mathbf{c}_j^{(i)}\|^2$$

$$\text{s.t. } 2\Re(\mathbf{z}_j^{(i)H} \mathbf{X}_{jk} \mathbf{c}_j^{(i)}) - \mathbf{z}_j^{(i)H} \mathbf{X}_{jk} \mathbf{z}_j^{(i)} \geq \eta_{jk} \ \forall k.$$

6:         Let $\mathbf{c}_j^{(i)}$ denote the optimal value obtained from $\widetilde{\mathcal{Q}}_j(\boldsymbol{\eta}_j, \mathbf{z}_j^{(i)})$, then set $\mathbf{z}_j^{(i+1)} \leftarrow \mathbf{c}_j^{(i)}$.
7:     **until** Convergence
8: **end for**
9: Compute the precoding vectors $\mathbf{w}_j = \mathbf{F}_j \mathbf{c}_j$ $\forall j \in \mathcal{G}$.

---

**Proposition 3.** *Algorithm 1 converges to a point satisfying the KKT conditions of* $\overline{\mathcal{Q}}(\boldsymbol{\eta})$ *, while providing a feasible solution for* $\mathcal{Q}(\boldsymbol{\eta})$ *.*

*Proof.* Please refer to the Appendix C. $\qquad \square$

Now let us consider $\overline{\mathcal{F}}(\boldsymbol{\eta}, P)$ and $\mathcal{F}(\boldsymbol{\eta}, P)$. A solution to these two problems can be achieved by first applying Algorithm 1 and then using Proposition 1. Besides, we can directly apply the SCA technique to $\overline{\mathcal{F}}(\boldsymbol{\eta}, P)$ and find a solution to these two problems, similar to Algorithm 1. The latter approach is presented in Algorithm 2 and we have the following proposition for Algorithm 2.

**Proposition 4.** *Algorithm 2 converges to a KKT point of* $\overline{\mathcal{F}}(\boldsymbol{\eta}, P)$ *, while providing a feasible solution to* $\mathcal{F}(\boldsymbol{\eta}, P)$ *.*

*Proof.* The proof follows the same lines as the proof of Proposition 3. $\qquad \square$

---

**Algorithm 2** The MMF BDZF-SCA Algorithm

1: Compute $\mathbf{F}_j$ $\forall j \in \mathcal{G}$.
2: Select an arbitrary set $\mathbf{z}^{(1)} := \{\mathbf{z}_j^{(1)}\}_{j=1}^{G}$ such that $\sum_{j=1}^{G} \|\mathbf{z}_j^{(1)}\|^2 \leq P$.
3: **repeat**
4:     Solve:

$$\widetilde{\mathcal{F}}(\boldsymbol{\eta}, P, \mathbf{z}^{(i)}) : \max_{\{\mathbf{c}_j^{(i)}\}} \min_j \min_k \frac{1}{\eta_{jk}} \Big[ 2\Re(\mathbf{z}_j^{(i)H} \mathbf{X}_{jk} \mathbf{c}_j^{(i)})$$
$$- \mathbf{z}_j^{(i)H} \mathbf{X}_{jk} \mathbf{z}_j^{(i)} \Big]$$

$$\text{s.t.} \quad \sum_{j=1}^{G} \|\mathbf{c}_j^{(i)}\|^2 \leq P.$$

5:     Let $\{\mathbf{c}_j^{(i)}\}_{j=1}^{G}$ denote the optimal value obtained from $\widetilde{\mathcal{F}}(\boldsymbol{\eta}, P, \mathbf{z}^{(i)})$, then $\forall j \in \mathcal{G}$ set $\mathbf{z}_j^{(i+1)} \leftarrow \mathbf{c}_j^{(i)}$.
6: **until** Convergence
7: Generate the precoding vectors $\mathbf{w}_j = \mathbf{F}_j \mathbf{c}_j$ $\forall j \in \mathcal{G}$.

---

### D. Computational Complexity

The computational load of Algorithms 1 and 2 is now assessed in terms of the number of required flops as follows. Note that both algorithms consist of three steps. The first step computes $\{\mathbf{F}_j; \forall j \in \mathcal{G}\}$ and requires $8N \sum_{j=1}^{G}(K - K_j)^2 - 8/3 \sum_{j=1}^{G}(K - K_j)^3$ flops using the QR-decomposition [33], [34]. The second step aims at designing the inner layer precoding vectors $\{\mathbf{c}_j; \forall j \in \mathcal{G}\}$ – as detailed in lines 2 to 8 (2 to 6) of Algorithm 1 (Algorithm 2). Since $\widetilde{\mathcal{Q}}(\boldsymbol{\eta}, \mathbf{z})$ and $\widetilde{\mathcal{F}}(\boldsymbol{\eta}, P, \mathbf{z})$ are both convex, they can be solved at each iteration using standard techniques with a worst case complexity of $\mathcal{O}(N^3)$ [27]. Therefore, the number of flops required by the second step is $\mathcal{O}(IN^3)$ with $I$ being the number of iterations required to converge. As it will be observed in Section V, only a few iterations are needed to reach a satisfying solution even for large $N$. The third step calculates the composite precoding vectors $\mathbf{w}_j = \mathbf{F}_j \mathbf{c}_j$ and requires $8GN^2 - 8(G-1)KN$ flops. In large-scale antenna array systems, i.e., where $N \gg K$, the overall complexity of the proposed algorithm is dominated by the second step and it is of $\mathcal{O}(N^3)$. Taking into account that the complexity of SDR based techniques is greater than $\mathcal{O}(N^6)$ [4], a reduction by a factor of $N^3$ is achieved through Algorithms 1 and 2.

## IV. A HEURISTIC INNER LAYER OF ORDER $\mathcal{O}(N)$

In the previous section, it was shown that the complexity of the proposed algorithms is of $\mathcal{O}(N^3)$, which is due to the application of SCA technique to find the inner layer precoding vectors, i.e., $\{\mathbf{c}_j\}_{j=1}^{G}$. Therefore, the inner layer retrieval may still be computationally expensive when $N$ is relatively large. Moreover, it requires optimization packages for solving the convex problems $\widetilde{\mathcal{Q}}(\boldsymbol{\eta}, \mathbf{z})$ and $\widetilde{\mathcal{F}}(\boldsymbol{\eta}, P, \mathbf{z})$, which may not be available on every hardware platform. Therefore, in what follows, we present a simple, yet effective, heuristic algorithm for computing the inner layer precoding vectors of $\overline{\mathcal{Q}}(\boldsymbol{\eta})$ with a complexity of $\mathcal{O}(N)$. Then, by employing Proposition 1 and the solution obtained for $\overline{\mathcal{Q}}(\boldsymbol{\eta})$, we compute an approximate solution for $\overline{\mathcal{F}}(\boldsymbol{\eta}, P)$. Therefore, the complexity of simultaneously finding an inner layer precoder for both problems becomes $\mathcal{O}(N)$.

The proposed heuristic algorithm aims at computing the precoding vector $\mathbf{c}_j \; \forall j \in \mathcal{G}$, in $K_j$ sequential steps. The algorithm has two main parts, the ordering part and the successive precoder design part. Assuming that the $K_j$ UEs in group $j$ are labeled from $1$ to $K_j$, the ordering part will re-label them by a bijective function $f_j : \{1, \ldots, K_j\} \rightarrow \{\mu_{j1}, \ldots, \mu_{jK_j}\}$, where $\mu_{jk} = f_j(i)$ means that the UE who was labeled as $i$ is now re-labeled as $\mu_{jk}$ and will be served in $k$th step of the algorithm, $k \in \{1, \ldots, K_j\}$. Therefore, the new labels, $\{\mu_{jk}\}_{k=1}^{K_j}$, will determine the order by which the UEs in group $j$ are served in each step. The successive precoder design part, designs the precoding vector of group $j$ in $K_j$ steps such that in $k$th step the requested SNR of UE $\mu_{jk}$ is met with minimum power consumption while the SNR of the previous $k-1$ ordered UEs, i.e., $\{\mu_{jt}\}_{t=1}^{k-1}$, is not violated. We will detail the successive precoder design and the user ordering in the following two subsections.

### A. The Successive Precoder

Assume that $\forall j \in \mathcal{G}$ the UE ordering is given, i.e., $\{\mu_{jk}\}_{k=1}^{K_j}$ is known. Denote by $\mathbf{c}_j^{(k)}$ the precoding vector $\mathbf{c}_j$ at $k$th step, then it is computed as follows:

$$\mathbf{c}_j^{(k)} = \mathbf{c}_j^{(k-1)} + \alpha_j^{(k)} \mathbf{d}_j^{(k)} \quad k \in \{1, \ldots, K_j\} \quad (30)$$

where $\mathbf{d}_j^{(k)} \in \mathbb{C}^{N-\tau_j}$ is a unit norm vector and $\alpha_j^{(k)} \in \mathbb{C}$. In what follows, we explain how $\mathbf{d}_j^{(k)}$ and $\alpha_j^{(k)}$ should be designed such that the SNR constraint of $\mu_{jk}$ is met with minimum power consumption while the SNR of $\{\mu_{jt}\}_{t=1}^{k-1}$ is not violated.

We start by initializing the precoding vector $\mathbf{c}_j^{(1)}$ of UE $\mu_{j1}$ such that its own SNR constraint, i.e., $|\overline{\mathbf{g}}_{j\mu_{j1}}^H \mathbf{c}_j^{(1)}|^2 \geq \eta_{j\mu_{j1}}$, is met with equality. This yields $\mathbf{c}_j^{(1)} = (\sqrt{\eta_{j\mu_{j1}}}/\|\overline{\mathbf{g}}_{j\mu_{j1}}\|^2)\overline{\mathbf{g}}_{j\mu_{j1}}$. For $k \in \{2, \ldots, K_j\}$, the vectors $\mathbf{d}_j^{(k)}$ must be chosen such that the previously satisfied $k-1$ SINR constraints are not violated. This is achieved by selecting $\mathbf{d}_j^{(k)}$ orthogonal to $\{\overline{\mathbf{g}}_{j\mu_{ji}}\}_{i=1}^{k-1}$, i.e., $\overline{\mathbf{g}}_{j\mu_{ji}}^H \mathbf{d}_j^{(k)} = 0$ for $i = 1, \ldots, k-1$. To this end, $\{\mathbf{d}_j^{(k)}\}_{k=2}^{K_j}$ are computed using the Gram–Schmidt procedure, which produces $\mathbf{d}_j^{(k)} = \mathbf{u}_j^{(k)}/\|\mathbf{u}_j^{(k)}\|$ with

$$\mathbf{u}_j^{(k)} = \overline{\mathbf{g}}_{j\mu_{jk}} - \sum_{i=1}^{k-1} \frac{\mathbf{u}_j^{(i)H}\overline{\mathbf{g}}_{j\mu_{jk}}}{\|\mathbf{u}_j^{(i)}\|^2}\mathbf{u}_j^{(i)} \quad (31)$$

being the component of $\overline{\mathbf{g}}_{j\mu_{jk}}$ orthogonal to the subspace spanned by $\{\mathbf{u}_j^{(i)}\}_{i=1}^{k-1}$. Once the unit norm vectors $\mathbf{d}_j^{(k)}$ are computed, we proceed with the design of coefficients $\{\alpha_j^{(k)}\}_{k=2}^{K_j}$. In particular, each $\alpha_j^{(k)}$ is chosen such that the power consumption in step $k$, given by $\|\mathbf{c}_j^{(k)}\|^2 = \|\mathbf{c}_j^{(k-1)}\|^2 + |\alpha_j^{(k)}|^2$, is minimized while satisfying the $k$th SINR constraint. More precisely, $\alpha_j^{(k)}$ must be computed as the solution of the following problem:

$$\min_{\alpha_j^{(k)}} \quad |\alpha_j^{(k)}|^2 \quad \text{s.t.} \quad |\overline{\mathbf{g}}_{j\mu_{jk}}^H \mathbf{c}_j^{(k)}|^2 \geq \eta_{j\mu_{jk}}. \quad (32)$$

As shown in the Appendix D (see also [12]), the optimal $\alpha_j^{(k)} = |\alpha_j^{(k)}| \exp(\mathrm{i} \angle \alpha_j^{(k)})$ is computed as:

$$\angle \alpha_j^{(k)} = -\angle \rho_j^{(k)} \quad (33)$$

$$|\alpha_j^{(k)}| = \frac{-|\rho_j^{(k)}| + \sqrt{|\rho_j^{(k)}|^2 - |\overline{\mathbf{g}}_{j\mu_{jk}}^H \mathbf{d}_j^{(k)}|^2 \left(|\overline{\mathbf{g}}_{j\mu_{jk}}^H \mathbf{c}_j^{(k-1)}|^2 - \eta_{j\mu_{jk}}\right)}}{|\overline{\mathbf{g}}_{j\mu_{jk}}^H \mathbf{d}_j^{(k)}|^2} \quad (34)$$

with $\rho_j^{(k)} = \overline{\mathbf{g}}_{j\mu_{jk}}^H \mathbf{d}_j^{(k)} \mathbf{c}_j^{(k-1)H} \overline{\mathbf{g}}_{j\mu_{jk}}$. In the sequel, the above results are used to sort the UEs according to a *worst-first* policy, which is observed to achieve close-to-optimal performance by means of numerical results in Section V.

### B. User Ordering

At this stage, we are only left with the computation of the UE ordering indices $\{\mu_{jk}\}$. A possible solution is illustrated in [12], for the QoS problem in single-group multicasting systems. More specifically, denote by $\mathcal{S}_j^{(k-1)} =$ $\{\mu_{j1}, \ldots, \mu_{j(k-1)}\}$ the set of indices of the ordered UEs at step $k-1$ and call $\mathcal{R}_j^{(k-1)}$ the set of indices of the remaining ones, i.e., $\mathcal{R}_j^{(k-1)} = \{1 \ldots, K_j\} \setminus \{f_j^{-1}(\mu_{jt})\}_{t=1}^{k-1}$. Then, in [12] the set $\mathcal{S}_j^{(k)}$ is computed as $\mathcal{S}_j^{(k)} = \mathcal{S}_j^{(k-1)} \cup \{\mu_{jk}\}$ with

$$\mu_{jk} = \underset{i \in \mathcal{R}_j^{(k-1)}}{\arg\min} \quad \frac{|\overline{\mathbf{g}}_{ji}^H \mathbf{c}_j^{(k-1)}|^2}{\eta_{ji}} \tag{35}$$

corresponding to the UE index in $\mathcal{R}_j^{(k-1)}$ that has the weakest ratio (or also the most violated constraint) between the provided SNR up to step $k$, given by $|\overline{\mathbf{g}}_{ji}^H \mathbf{c}_j^{(k-1)}|^2$, and the requested one given by $\eta_{ji}$. The above procedure has the following two drawbacks. Firstly, it needs to calculate at each stage $k$ the quantities $|\overline{\mathbf{g}}_{ji}^H \mathbf{c}_j^{(k-1)}|^2 \ \forall i \in \mathcal{R}_j^{(k-1)}$, which requires $\mathcal{O}(K_j^2 N)$ flops for group $j$. This is costly if $N$ and $K_j$ are large. Secondly, it does not take into account the extra amount of power $|\alpha_j^{(k)}|^2$ required at stage $k$ to meet the SNR constraint of the selected UE. To see how this comes about, consider a generic UE $i \in \mathcal{R}_j^{(k-1)}$ such that at stage $k$ the ratio $|\overline{\mathbf{g}}_{ji}^H \mathbf{c}_j^{(k-1)}|^2 / \eta_{ji}$ takes a very high value. This might happen, for example, because its own channel vector $\overline{\mathbf{g}}_{ji}$ is almost collinear to the channel vectors of the UEs selected in the previous $k-1$ stages. According to (35), such a UE will be selected at the very end of the procedure. This, however, would result in a huge power consumption because the Gram–Schmidt procedure will only have a restricted number of degrees of freedom to make $\mathbf{c}_j^{(K_j)}$ orthogonal to $\overline{\mathbf{g}}_{ji}$ for $i = 1, \ldots, K_j - 1$ and at the same time to meet the requested SINR. In other words, the procedure in (35) sorts the UEs according to a *best-first* criterion such that higher priority is given to the UEs requiring low power to meet their SNR constraints.

Unlike [12], we make use of the power increase (34) at each stage $k$ to order the UEs within each group according to a *worst-first* criterion. As mentioned before, this choice is motivated by the fact that the Gram–Schmidt procedure in (31) progressively reduces the available degrees of freedom as $k$ tends to $K_j$. Therefore, since power consumption is dominated by UEs with the worst conditions (according to some criterion), higher priority should be given to these UEs. Mathematically, we propose to compute the index $\mu_{jk}$ at step $k$ as follows:

$$\mu_{jk} = \underset{i \in \mathcal{R}_j^{(k-1)}}{\arg\max} |\alpha_{ji}^{(k)}|^2 \tag{36}$$

with

$$|\alpha_{ji}^{(k)}| = \frac{-|\rho_{ji}^{(k)}| + \sqrt{|\rho_{ji}^{(k)}|^2 - |\overline{\mathbf{g}}_{ji}^H \mathbf{d}_j^{(k)}|^2 (|\overline{\mathbf{g}}_{ji}^H \mathbf{c}_j^{(k-1)}|^2 - \eta_{ji})}}{|\overline{\mathbf{g}}_{ji}^H \mathbf{d}_j^{(k)}|^2} \tag{37}$$

and $\rho_{ji}^{(k)} = \overline{\mathbf{g}}_{ji}^H \mathbf{d}_j^{(k)} \mathbf{c}_j^{(k-1)H} \overline{\mathbf{g}}_{ji}$. As seen, $\mu_{jk}$ corresponds to the UE index in $\mathcal{R}_j^{(k-1)}$ for which the incremental power $|\alpha_{ji}^{(k)}|^2$ at stage $k$ takes the maximum value. Note that the computational cost of this operation is still $\mathcal{O}(K_j^2 N)$ flops as for [12]. To further reduce the computational burden, we propose an alternative approach that exploits the inherent characteristic of large-scale antenna systems. As $N$ is large, each user

$i \in \mathcal{R}_j^{(k-1)}$ can use the excess degree of freedom, provided by the large number of antennas, to chose $\mathbf{d}_j^{(k)}$ as collinear as possible to $\overline{\mathbf{g}}_{ji}$ while almost nulling the interference generated to the other UEs, i.e., $|\overline{\mathbf{g}}_{ji}^H \mathbf{c}_j^{(k-1)}| \approx 0$. Therefore, by replacing $|\overline{\mathbf{g}}_{ji}^H \mathbf{d}_j^{(k)}|^2$ with $\|\overline{\mathbf{g}}_{ji}\|^2$ and neglecting the term $|\overline{\mathbf{g}}_{ji}^H \mathbf{c}_j^{(k-1)}|$, the right-hand-side of (37) reduces to $\frac{\eta_{ji}}{\|\overline{\mathbf{g}}_{ji}\|^2}$. This means that UEs in group $j$ can be ordered by simply sorting the following ratios in a descending order:

$$\left\{\frac{\eta_{j1}}{\|\overline{\mathbf{g}}_{j1}\|^2}, \ldots, \frac{\eta_{jK_j}}{\|\overline{\mathbf{g}}_{jK_j}\|^2}\right\} \quad \forall j \in \mathcal{G}. \tag{38}$$

In other words, higher priority should be given to those UEs that have bad channel conditions compared to the target SNRs. In doing so, no greedy strategy is required for UE ordering, thereby reducing the total number of flops to $\mathcal{O}(K_j N)$. Based on the above discussion, a heuristic solution is proposed in Algorithm 3 for the inner layer. Numerical results are used in Section V to make comparisons among the above ordering policies in different settings. As it will be seen, the ordering policy of (38) largely outperforms the strategy of [12].

### C. The Proposed Heuristic Inner Layer Precoder

Collecting the results achieved in Sections IV-A and IV-B, we present the following heuristic algorithm to design the inner layer precoder of $\overline{\mathcal{Q}}(\boldsymbol{\eta})$. To emphasis on the simplicity of Algorithm 3 and to enable the reproducibility of our results, its MATLAB code is provided in [35].

---

**Algorithm 3** A heuristic algorithm of the inner layer for solving $\overline{\mathcal{Q}}(\boldsymbol{\eta})$

---

1: **for** $j = 1$ to $G$ **do**
2:     Sort the UEs in group $j$ in descending order based on $\{\frac{\eta_{ji}}{\|\overline{\mathbf{g}}_{ji}\|^2}\}$ and label the list as $\{\mu_{j1}, \ldots, \mu_{jK_j}\}$, respectively.
3:     Compute $\{\mathbf{d}_j^{(k)}\}_{k=1}^{K_j}$ using the Gram–Schmidt procedure in (31).
4:     Set $\mathbf{c}_j^{(1)} = \frac{\sqrt{\eta_{j1}}}{\|\overline{\mathbf{g}}_{j\mu_{j1}}\|^2} \mathbf{g}_{j\mu_{j1}}$.
5:     **for** $k = 2$ to $K_j$ **do**
6:         **if** $|\overline{\mathbf{g}}_{j\mu_{jk}}^H \mathbf{c}_j^{(k-1)}|^2 < \eta_{j\mu_{jk}}$ **then**
7:             Compute $\alpha_j^{(k)}$ through (33) and (34).
8:             Update $\mathbf{c}_j^{(k)} = \mathbf{c}_j^{(k-1)} + \alpha_j^{(k)} \mathbf{d}_j^{(k)}$.
9:         **end if**
10:     **end for**
11: **end for**

---

The complexity of Algorithm 3 can be evaluated as follows. Observe that evaluating the terms $\{\eta_{ji}/\|\overline{\mathbf{g}}_{ji}\|^2\}$ for group $j$ requires $4(N - \tau_j)$ flops whereas sorting a list of size $K_j$ needs $\mathcal{O}(K_j \log(K_j))$ flops. Therefore, the flop counts for UE ordering in line 2 is $4K_j(N - \tau_j) + \mathcal{O}(K_j \log(K_j))$. The Gram–Schmidt procedure of line 4 can be performed through the QR decomposition, which requires $8(N - \tau_j)K_j^2 - \frac{8}{3}K_j^3$ flops [33]. The computation of $\mathbf{c}_j^{(1)}$ needs $2(N - \tau_j + 1)$ flops. The condition $|\overline{\mathbf{g}}_{j\mu_{jk}}^H \mathbf{c}_j^{(k-1)}|^2 < \eta_{jk}$ in line 6 avoids to waste power for those UEs whose requested SNR constraints are

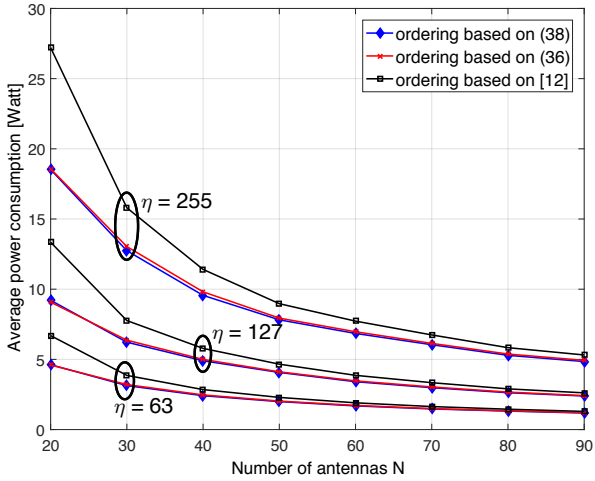

Fig. 1: Average power consumption of the QoS problem, comparing different ordering policies for $G = 1$ and $K = 20$.

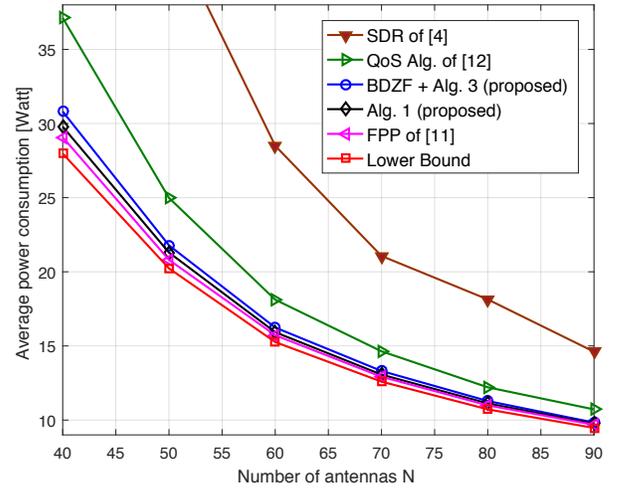

Fig. 2: Average power consumption of the QoS problem with $\eta_{jk} = 255 \ \forall j, k$.

already met (more details on this are given in the Appendix). Lines 6 to 10 require $\mathcal{O}(N - \tau_j)$ flops, and as the condition of line 6 is true at most $K_j - 1$ times, the flops required by lines 5 to 11 is $\mathcal{O}(K_j(N - \tau_j))$. Summing all the above terms together, the complexity of Algorithm 3 is found to be $\mathcal{O}(N)$, thereby reducing the complexity of the inner layer precoder by a factor of $N^2$.

Note that by jointly employing Proposition 1 and Algorithm 3, the approximated precoding vectors for $\overline{\mathcal{F}}(\boldsymbol{\eta}, P)$ can be computed as

$$\mathbf{c}^{\circ}_{j,\mathrm{BDZF-HEU}}(\boldsymbol{\eta}, P) = \sqrt{\frac{P}{P^{\star}_{\mathrm{BDZF-HEU}}(\boldsymbol{\eta})}} \mathbf{c}^{\star}_{j,\mathrm{BDZF-HEU}}(\boldsymbol{\eta}) \quad (39)$$

where $\{\mathbf{c}^{\star}_{j,\mathrm{BDZF-HEU}}(\boldsymbol{\eta})\}$ and $P^{\star}_{\mathrm{BDZF-HEU}}(\boldsymbol{\eta})$ denotes the precoding vectors and the total power consumption as obtained with Algorithm 3. Therefore the precoding vectors for $\mathcal{F}(\boldsymbol{\eta}, P)$ are given by $\{\mathbf{F}_j \mathbf{c}^{\circ}_{j,\mathrm{BDZF-HEU}}(\boldsymbol{\eta}, P)\}$.

## V. NUMERICAL RESULTS

Monte Carlo simulations are used to assess the performance of the proposed algorithms and to make comparisons with existing alternatives. In particular, we consider the algorithm presented in [4], which employs the SDR technique followed by a randomization and multicast multigroup power control (MMPC) policy.[3] Comparisons are also made with the asymptotic results of [23], the FPP based algorithm presented in [11], and the heuristic algorithms developed in [12]. A single-cell system with radius of 900 meters is considered with UEs being distributed uniformly and randomly in the cell excluding an inner circular area of radius 100 meters. For each value of $N$, the average values of power consumption or minimum SINR of the system are obtained from 100 different channel realizations and UE distributions. We assume (if not otherwise specified) that there are $G = 3$ multicasting groups, each counting $K_j = 10$ UEs (such that $K = 30$). The channel vector $\mathbf{g}_{jk}$ between UE $k$ in group $j$ and the BS is modeled as

$\mathbf{g}_{jk} = \sqrt{\beta_{jk}} \mathbf{h}_{jk}$ where $\mathbf{h}_{jk} \sim \mathcal{CN}(\mathbf{0}, \mathbf{I}_N)$ represents the small scale fading and $\beta_{jk}$ accounts for the large scale attenuation given by $\beta_{jk} = -128.1 - 37.6 \log_{10} d_{jk}$ dB with $d_{jk}$ being the distance between the UE and the BS expressed in kilometers [36]. The noise power spectral density is assumed to be $-174$ dBm/Hz, and the channel bandwidth is 20 MHz [28]. All the simulations are performed on a 64-bit Linux operating system with Intel Xeon processor E5-1680 v3.

Fig. 1 compares the average power consumption of the ordering policies proposed in [12] with those given by (36) and (38), for $G = 1$, $K = 20$ and $\eta = 63$, 127, and 255 (which correspond to a target rate for each UE of 6, 7 and 8 bit/s/Hz, respectively). The proposed ordering policies are seen to outperform the ordering of [12]. Moreover, the simple ordering policy of (38) has even a slightly better performance than (36). Note that, as the ordering belongs to the heuristic inner layer of the proposed precoder and as the outer layer removes the effect of inter-group interference, the same conclusion holds for $G > 1$. Based on the above results, the simpler ordering policy presented in (38) will be used in the remainder of this section.

Fig. 2 depicts the average power consumption of the QoS problem versus the number of antennas $N$ at the BS. We assume that $\eta_{jk} = 255$ for all UEs (corresponding to 8 bit/s/Hz/UE), and it is chosen in agreement with 5G rate requirements [37], but the conclusions generically hold for all other values of $\boldsymbol{\eta}$. The performance of Algorithm 1, and the combination of BDZF and Algorithm 3 is compared to other existing algorithms. As the QoS problem is NP-hard, a lower bound of the problem is also presented as a benchmark [4]. Observe that, the proposed algorithms outperform the SDR-based solution in [4] and the heuristic one in [12], while they have nearly the same performance as [11]. However, this is achieved at a much lower complexity and computational cost as detailed next. Note that for $N \geq 60$ both algorithms are at most 6% away from the lower bound and this gap reduces as $N$ grows large, while for SDR technique this gap is 87% and reduces slowly by adding more antennas.

Fig. 3 illustrates the average minimum SINR of the MMF

---

[3]For the randomization phase, 100 samples are generated using the Gaussian randomization method [4].

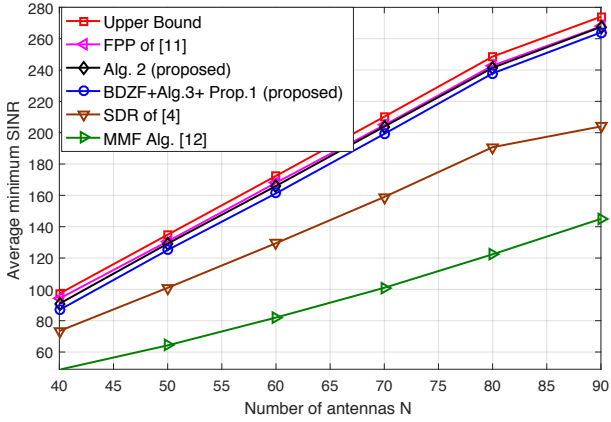

Fig. 3: Average minimum SINR of the MMF problem for $P = 10$ Watt.

TABLE I: The average time (in seconds) required to solve QoS, MMF, or both of them.

| | QoS Problem | | MMF Problem | | Both QoS and MMF | |
| --- | --- | --- | --- | --- | --- | --- |
| | SDR[4] | FPP[11] | SDR[4] | FPP[11] | Alg.1+Pr.1 | BDZF+Alg.3+Pr.1 |
| $N=40$ | 55 | 41 | 419 | 356 | 11.3 | $2.5 \times 10^{-3}$ |
| $N=50$ | 67 | 51 | 579 | 450 | 11.6 | $2.8 \times 10^{-3}$ |
| $N=60$ | 84 | 61 | 798 | 507 | 11.7 | $3.1 \times 10^{-3}$ |
| $N=70$ | 110 | 75 | 1151 | 617 | 11.9 | $3.5 \times 10^{-3}$ |
| $N=80$ | 146 | 87 | 1549 | 727 | 12.2 | $4.0 \times 10^{-3}$ |
| $N=90$ | 182 | 107 | 2050 | 865 | 12.5 | $4.5 \times 10^{-3}$ |

problem versus $N$. The available power at the BS is considered to be 10 Watt. In this figure, the performance achieved by Algorithm 2, and the combination of BDZF, Algorithm 3, and Proposition 1 is compared to other existing algorithms. Also, the upper bound of the problem is depicted as a benchmark. Similar to the results of Fig. 2, the proposed algorithms largely outperform [4] and [12], while nearly having the same performance as [11]. However, this is achieved for a computational cost that is significantly smaller than other algorithms as detailed next. Observe that Algorithm 2 is within 5% of the upper bound with just $N = 50$ antennas. Also, Algorithm 3 jointly with BDZF and Proposition 1 achieve the same target with $N = 70$ antennas.

To assess the computational complexity of the investigated algorithms more intuitively, beside the complexity analysis of Section III-D and Section IV-C, we also present the computation time required to approximately solve $\mathcal{Q}(\boldsymbol{\eta})$ and $\mathcal{F}(\boldsymbol{\eta}, P)$ versus $N$, in Table I. The table presents the average time (in seconds) required to solve the QoS and MMF problems. The second and third columns report the average time required by the SDR and the FPP algorithms to solve an instance of the QoS problem. The fourth and fifth columns present the average required time by the same algorithms to solve an instance of the MMF problem. Note the increase in time from the QoS problem to the MMF problem, as in the SDR and FPP algorithms the MMF is solved by iteratively applying the QoS algorithm. The sixth and seventh columns of the table present the average time required to solve both QoS and MMF problems simultaneously using the proposed algorithms.

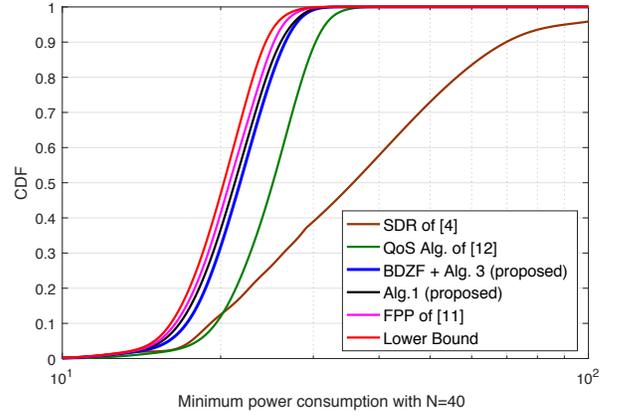

(a) $N = 40$.

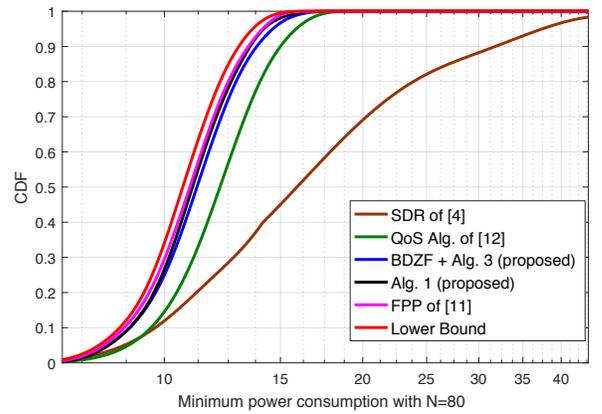

(b) $N = 80$.

Fig. 4: CDF of minimum power consumption (QoS problem) of the system.

Note that not only we solve both problems at the same time with good performance, but also the required time has reduced significantly. As an example, for $N = 90$, the combination of BDZF, Algorithm 3, and Proposition 1, solves both problems in less than 5 milliseconds, while the SDR and the FPP algorithms require 2050 or 865 seconds, respectively, just to solve the MMF problem. At the same time, as shown in Figs. 2 and 3, the solution provided by joint application of the BDZF, Algorithm 3, and Proposition 1 is nearly as good as the solution achieved from the FPP Algorithm, and significantly outperforms the SDR Algorithm.

Figs. 4 and 5 present the cumulative distribution function (CDF) of the precoder power consumption and the minimum SINR of the system for the QoS and MMF problems, respectively. For the QoS problem, the requested SINR by each user is assumed to be $255$, and for the MMF problem the available power at the BS is considered to be 10 Watt. Unlike Figs. 2 and 3 that provide the average of minimum power consumption or the average of minimum SINR of the system, these figures provide a clear vision on the distribution of these quantities, for the existing and proposed algorithms. It is seen that, as we increase $N$ from 40 to 80, the CDF curves of our proposed algorithms become closer to the optimal bound, and also improve significantly in terms of performance, thanks to the large number of antennas. As an example, for the QoS problem with $N = 40$, the power consumption is greater than 20 Watt 60% of the times, while with $N = 80$ our proposed algorithms always meet the requested SINRs with less than 17 Watt. Also, for the MMF problem none of the algorithms can provide a minimum SINR bigger than 200 with $N = 40$, while for $N = 80$ our proposed algorithms can provide a minimum SINR bigger than 200 in 80% of the times. Fig. 5 also contains the CDF of the asymptotic approach of [23]. Notice that the asymptotic approach can never provide an SINR which is bigger than 20 (or 100) with $N = 40$ (or $N = 80$) antennas and its insufficiency is detailed in [23].

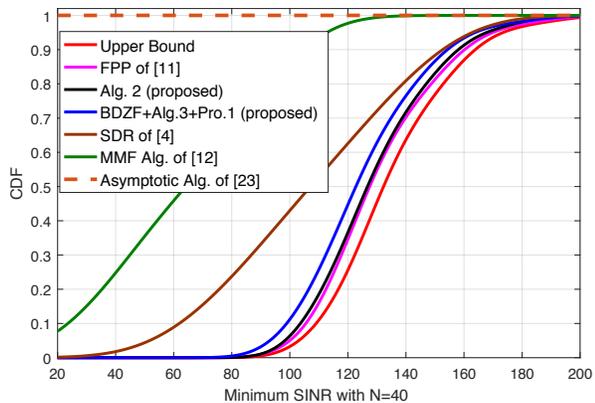

(a) $N = 40$.

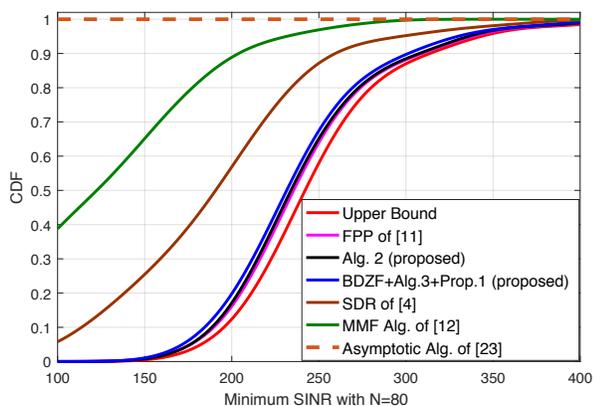

(b) $N = 80$.

Fig. 5: CDF of minimum SINR (MMF problem) of the system with $P = 10$.

In Section III, we have elaborated the BDZF-SCA approach and proved the convergence of Algorithms 1 and 2, but we have not specified the number of iterations required by each algorithm to converge. Table II presents the average number of iterations required by Algorithms 1 and 2 to achieve convergence for different values of $G, K$ and $N$. Denoting the objective value achieved at the $k$th iteration of either Algorithm 1 or 2 as $\varepsilon(k)$, the convergence condition of Table

TABLE II: Average number of iterations required by Algorithms 1 and 2.

| | $G = 2, K = 10$ | | $G = 3, K = 10$ | | $G = 3, K = 15$ | |
|---|---|---|---|---|---|---|
| | Alg. 2 | Alg. 1 | Alg. 2 | Alg. 1 | Alg. 2 | Alg. 1 |
| $N = 40$ | 15.12 | 16.56 | 15.38 | 16.90 | 15.95 | 17.00 |
| $N = 50$ | 15.14 | 16.61 | 15.50 | 16.94 | 16.12 | 17.11 |
| $N = 60$ | 15.21 | 16.74 | 15.54 | 16.97 | 16.23 | 17.14 |
| $N = 70$ | 15.40 | 16.88 | 15.60 | 17.04 | 16.46 | 17.40 |
| $N = 80$ | 15.44 | 16.94 | 15.60 | 17.08 | 16.52 | 17.51 |
| $N = 90$ | 15.53 | 17.10 | 15.62 | 17.12 | 16.68 | 17.56 |

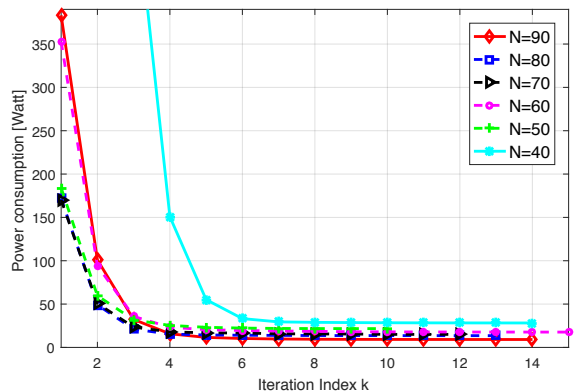

(a) Algorithm 1.

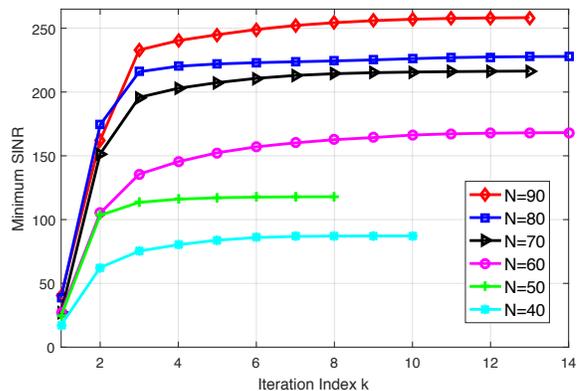

(b) Algorithm 2.

Fig. 6: The convergence behavior of Algorithms 1 and 2 for different number of antennas $N$.

II is $\frac{|\varepsilon(k+1) - \varepsilon(k)|}{\varepsilon(k)} < 10^{-3}$. Also, Fig. 6 illustrates $\varepsilon(k)$ for both algorithms at each iteration index $k$ for different number of antennas $N$. As it is seen, both algorithms converge in a few iterations for any value of $N$.

So far, we have assumed that the BS has perfect knowledge of the channel vectors $\{\mathbf{g}_{jk}\}$. Although some of the analysis can in principle be extended (to some extent) to the scenario with imperfect CSI, we believe that this is out of the scope of this work and thus it is left for the future. To partially fulfill this lack, we now investigate the impact of imperfect CSI on the performance of the proposed algorithms. A time-division-protocol (TDD) is employed such that channel estimation can be performed in the uplink on the basis of UE pilot signals and then used in the downlink. We assume that pilots of length $\tau_p = K$ are used, with power equal to 1 Watt. The estimates

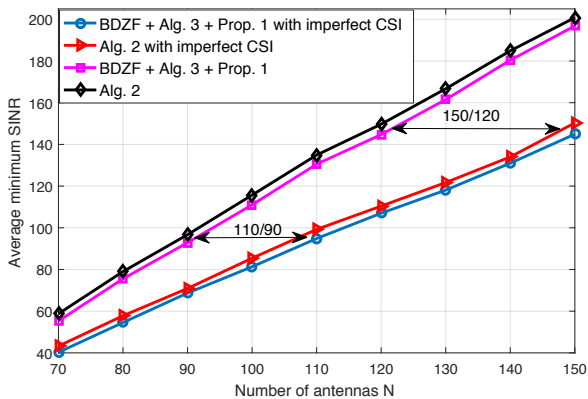

Fig. 7: Evaluating the impact of imperfect CSI at the BS for the MMF Problem with $G=4$, $K_j=15$ $\forall j$ and $P=10$ Watt.

of channel vectors $\{\mathbf{g}_{jk}\}$ are computed at the BS using an MMSE estimator. This yields [38]

$$\hat{\mathbf{g}}_{jk} = \frac{\sqrt{\tau_p}\beta_{jk}}{\sigma^2 + \tau_p\beta_{jk}}\left(\sqrt{\tau_p}\mathbf{g}_{jk} + \mathbf{n}\right) \quad \forall j,k \quad (40)$$

where $\mathbf{n} \sim \mathcal{CN}(\mathbf{0}, \sigma^2\mathbf{I}_N)$ is the additive Gaussian noise. Fig. 7 reports the performance of the proposed algorithms for MMF when perfect and imperfect CSI (as given in (40)) is available at the BS. We assume that $G = 4$ and $K_j = 15$ for $j = 1, \ldots, 4$. As expected, imperfect CSI degrades the performance of the proposed algorithms. However, such a performance loss can be compensated by using more antennas at the BS. Quantitively speaking, with imperfect CSI $N$ must be roughly increased by a factor of $15\% - 30\%$ compared to the perfect CSI case. The higher $N$, the larger the factor. For example, to achieve the same performance of the perfect CSI case with $N = 90$ and 120, then 110 and 150 antennas are respectively needed with imperfect CSI, corresponding to a 22% and 25% increase.

## VI. Conclusions

Multicasting is an efficient technology to transmit distinct common data streams to multiple groups of users. The existing multicasting algorithms are either computationally expensive or exhibit poor performance when applied to large-scale systems with hundreds of antennas, as envisioned in next generation of wireless systems. In this paper, we designed new algorithms, which are tailored for physical layer multicasting in large-scale antenna systems. The proposed algorithms achieve good performance and are characterized by affordable computational complexity. This was achieved by using the large number of antennas to first cancel the intergroup interference and then reformulate both the QoS and MMF problems in simple forms. Two efficient algorithms for solving the simplified problems were presented. Unlike baseline methods that solve the MMF problem by iteratively solving the QoS problem, we showed how to solve both simultaneously with no extra cost.

## Appendix A - Proof of Proposition 1

Starting with the power constraint of (13) and replacing $\mathbf{c}_j$ with $\sqrt{\frac{P}{P^\star_{\mathrm{app}}(\boldsymbol{\eta})}}\mathbf{c}^\star_{j,\mathrm{app}}(\boldsymbol{\eta})$ we have

$$\sum_{j=1}^{G}\|\sqrt{\frac{P}{P^\star_{\mathrm{app}}(\boldsymbol{\eta})}}\mathbf{c}^\star_{j,\mathrm{app}}(\boldsymbol{\eta})\|^2 = \frac{P}{P^\star_{\mathrm{app}}(\boldsymbol{\eta})}\sum_{j=1}^{G}\|\mathbf{c}^\star_{j,\mathrm{app}}(\boldsymbol{\eta})\|^2 = P$$

which proves the feasibility of the proposed solution. For the achievable objective value of $\overline{\mathcal{F}}(\boldsymbol{\eta}, P)$ (or $\mathcal{F}(\boldsymbol{\eta}, P)$) using $\{\sqrt{\frac{P}{P^\star_{\mathrm{app}}(\boldsymbol{\eta})}}\mathbf{c}^\star_{j,\mathrm{app}}(\boldsymbol{\eta})\}$, we have

$$t^\circ_{app}(\boldsymbol{\eta}, P) = \min_j\min_k\frac{1}{\eta_{jk}}|\overline{\mathbf{g}}^H_{jk}\sqrt{\frac{P}{P^\star_{\mathrm{app}}(\boldsymbol{\eta})}}\mathbf{c}^\star_{j,\mathrm{app}}(\boldsymbol{\eta})|^2$$
$$= \frac{P}{P^\star_{\mathrm{app}}(\boldsymbol{\eta})}\min_j\min_k\frac{1}{\eta_{jk}}|\overline{\mathbf{g}}^H_{jk}\mathbf{c}^\star_{j,\mathrm{app}}(\boldsymbol{\eta})|^2.$$

Denote $\lambda := \min_j\min_k\frac{1}{\eta_{jk}}|\overline{\mathbf{g}}^H_{jk}\mathbf{c}^\star_{j,\mathrm{app}}(\boldsymbol{\eta})|^2$. As $\{\mathbf{c}^\star_{j,\mathrm{app}}(\boldsymbol{\eta})\}$ is a set of the precoding vectors of $\overline{\mathcal{Q}}(\boldsymbol{\eta})$, $\lambda \geq 1$. Therefore

$$t^\circ_{app}(\boldsymbol{\eta}, P) = \frac{P}{\frac{1}{\lambda}P^\star_{\mathrm{app}}(\boldsymbol{\eta})} \geq \frac{P}{P^\star_{\mathrm{app}}(\boldsymbol{\eta})}.$$

As $\frac{1}{\lambda}P^\star_{\mathrm{app}}(\boldsymbol{\eta})$ is an objective value of $\overline{\mathcal{Q}}(\boldsymbol{\eta})$ that can be achieved by $\{\frac{1}{\sqrt{\lambda}}\mathbf{c}^\star_{j,\mathrm{app}}(\boldsymbol{\eta})\}$, it is bigger than or equal to the optimal objective value of $\overline{\mathcal{Q}}(\boldsymbol{\eta})$, i.e., $\frac{1}{\lambda}P^\star_{\mathrm{app}}(\boldsymbol{\eta}) \geq P^\star(\boldsymbol{\eta})$ , and we have $P/P^\star_{\mathrm{app}}(\boldsymbol{\eta}) \leq t^\circ_{app}(\boldsymbol{\eta}, P) \leq P/P^\star(\boldsymbol{\eta})$.

## Appendix B - Proof of Proposition 2

Starting with the SINR constraint of (11) and replacing $\mathbf{c}_j$ with $\frac{1}{\sqrt{t^\circ_{app}(\boldsymbol{\eta}, P)}}\mathbf{c}^\circ_{j,\mathrm{app}}(\boldsymbol{\eta}, P)$, we have

$$\frac{|\overline{\mathbf{g}}^H_{jk}\mathbf{c}^\circ_{app}(\boldsymbol{\eta}, P)|^2}{t^\circ_{app}(\boldsymbol{\eta}, P)} = \frac{\eta_{jk}}{t^\circ_{app}(\boldsymbol{\eta}, P)}\frac{|\overline{\mathbf{g}}^H_{jk}\mathbf{c}^\circ_{j,\mathrm{app}}(\boldsymbol{\eta}, P)|^2}{\eta_{jk}} \overset{(a)}{\geq} \eta_{jk}$$

where in (a) is due to the fact that $t^\circ_{app}(\boldsymbol{\eta}, P)$ is the minimum weighted SINR among all UEs. Therefore $\{\frac{1}{\sqrt{t^\circ_{app}(\boldsymbol{\eta}, P)}}\mathbf{c}^\circ_{j,\mathrm{app}}(\boldsymbol{\eta}, P)\}$ is a feasible answer of $\overline{\mathcal{Q}}(\boldsymbol{\eta})$ and $\mathcal{Q}(\boldsymbol{\eta})$. For the objective value we have

$$P^\star_{\mathrm{app}}(\boldsymbol{\eta}) = \sum_{j=1}^{G}\|\frac{\mathbf{c}^\circ_{j,\mathrm{app}}(\boldsymbol{\eta}, P)}{\sqrt{t^\circ_{app}(\boldsymbol{\eta}, P)}}\|^2 \overset{(a)}{=} \frac{\sum_{j=1}^{G}\|\mathbf{c}^\circ_{j,\mathrm{app}}(\boldsymbol{\eta}, P)\|^2}{t^\circ_{app}(\boldsymbol{\eta}, P)} \leq \frac{P}{t^\circ_{app}(\boldsymbol{\eta}, P)}.$$

Denote $\lambda := \frac{1}{P}\sum_{j=1}^{G}\|\mathbf{c}^\circ_{j,\mathrm{app}}(\boldsymbol{\eta}, P)\|^2$. As $\{\mathbf{c}^\circ_{j,\mathrm{app}}(\boldsymbol{\eta}, P)\}$ is a set of precoding vectors of $\overline{\mathcal{F}}(\boldsymbol{\eta}, P)$, $\lambda \leq 1$. Therefore,

$$P^\star_{\mathrm{app}}(\boldsymbol{\eta}) = \frac{P}{\frac{1}{\lambda}t^\circ_{app}(\boldsymbol{\eta}, P)} \leq \frac{P}{t^\circ_{app}(\boldsymbol{\eta}, P)}.$$

Since $\frac{1}{\lambda}t^\circ_{app}(\boldsymbol{\eta}, P)$ is an objective value of $\overline{\mathcal{F}}(\boldsymbol{\eta}, P)$ achieved by $\{\frac{1}{\sqrt{\lambda}}\mathbf{c}^\circ_{j,\mathrm{app}}(\boldsymbol{\eta}, P)\}$, it is less than or equal to the optimal objective value of $\overline{\mathcal{F}}(\boldsymbol{\eta}, P)$, i.e., $\frac{1}{\lambda}t^\circ_{app}(\boldsymbol{\eta}, P) \leq t^\circ(\boldsymbol{\eta}, P)$. Hence we have $P/t^\circ(\boldsymbol{\eta}, P) \leq P^\star_{\mathrm{app}}(\boldsymbol{\eta}) \leq P/t^\circ_{app}(\boldsymbol{\eta}, P)$.

## Appendix C - Proof of Proposition 3

As $\mathbf{z}_j^{(1)H}\mathbf{X}_{jk}\mathbf{z}_j^{(1)} \geq \eta_{jk}\ \forall k \in \mathcal{K}_j, j \in \mathcal{G}$, $\mathbf{z}_j^{(1)}$ is a feasible solution of $\widetilde{\mathcal{Q}}_j^{(1)}(\boldsymbol{\eta}_j, \mathbf{z}_j^{(1)})$. Now consider the $(i+1)$th iteration of the problem $\forall i \in \{0, 1, \ldots\}$. $\forall k \in \mathcal{K}_j, j \in \mathcal{G}$ we have

$$2\Re\mathfrak{e}(\mathbf{z}_j^{(i+1)H}\mathbf{X}_{jk}\mathbf{c}_j^{(i+1)}) - \mathbf{z}_j^{(i+1)H}\mathbf{X}_{jk}\mathbf{z}_j^{(i+1)} \overset{a}{=} \\ 2\Re\mathfrak{e}(\mathbf{c}_j^{(i)H}\mathbf{X}_{jk}\mathbf{c}_j^{(i+1)}) - \mathbf{c}_j^{(i)H}\mathbf{X}_{jk}\mathbf{c}_j^{(i)} \quad (41)$$

where $(a)$ is due to our update rule, $\mathbf{z}_j^{(i+1)} \leftarrow \mathbf{c}_j^{(i)}$. Now if we set $\mathbf{c}_j^{(i+1)} = \mathbf{c}_j^{(i)}$, (41) reduces to $\mathbf{c}_j^{(i)H}\mathbf{X}_{jk}\mathbf{c}_j^{(i)}$ which is bigger than $\eta_{jk}$ due to (26) and (27). Therefore $\mathbf{c}_j^{(i)}$ is a feasible solution of $\widetilde{\mathcal{Q}}_j^{(i+1)}(\boldsymbol{\eta}_j, \mathbf{z}_j^{(i)})$. Hence the objective function of $(i+1)$th iteration is less than or equal to the objective function of $(i)$th iteration. As the objective function is bounded from below, by successively solving the problem we achieve a non-increasing bounded sequence. Therefore the algorithm converges. Due to (26), any internal precoding vector $\mathbf{c}_j$ that satisfies (27), will also satisfy (11) and as a result, any answer to $\widetilde{\mathcal{Q}}(\boldsymbol{\eta}, \mathbf{z})$ is a feasible answer to $\overline{\mathcal{Q}}(\boldsymbol{\eta})$ and therefore $\mathcal{Q}(\boldsymbol{\eta})$. Due to the update rule and the inner approximation in (27), the convergence point satisfies the KKT conditions for $\overline{\mathcal{Q}}(\boldsymbol{\eta})$ as detailed in [30].

## Appendix D - Solution to (32)

Hereby we prove the solution of (32), i.e., $\alpha_j^{(k)}$, is given by (33) and (34). We start with the SNR constraint $|\bar{\mathbf{g}}_{j\mu_{jk}}^H\mathbf{c}_j^{(k)}|^2 \geq \eta_{j\mu_{jk}}$, and replace $\mathbf{c}_j^{(k)}$ with $\mathbf{c}_j^{(k-1)} + \alpha_j^{(k)}\mathbf{d}_j^{(k)}$ using (30). Denote $|\bar{\mathbf{g}}_{j\mu_{jk}}^H\mathbf{d}_j^{(k)}|^2$, $2\,\mathrm{Re}(e^{j\angle\alpha_j^{(k)}}\,\bar{\mathbf{g}}_{j\mu_{jk}}^H\mathbf{d}_j^{(k)}\mathbf{c}_j^{(k-1)H}\bar{\mathbf{g}}_{j\mu_{jk}})$, and $|\bar{\mathbf{g}}_{j\mu_{jk}}^H\mathbf{c}_j^{(k-1)}|^2 - \eta_{j\mu_{jk}}$ as $A$, $B$, and $C$, respectively. The SNR constraint can be represented as

$$|\bar{\mathbf{g}}_{j\mu_{jk}}^H\mathbf{c}_j^{(k)}|^2 - \eta_{j\mu_{jk}} = A|\alpha_j^{(k)}|^2 + B|\alpha_j^{(k)}| + C \geq 0 \quad (42)$$

Notice that if $|\bar{\mathbf{g}}_{j\mu_{jk}}^H\mathbf{c}_j^{(k)}|^2 \geq \eta_{j\mu_{jk}}$, to minimize the power, no transmission shall be arranged for user $\mu_{jk}$, i.e., $\alpha_j^{(k)} = 0$, and the next user shall be served. Otherwise, $C < 0$. Now we transform (42) to an equality by introducing $\lambda \geq 0$ as follows

$$A|\alpha_j^{(k)}|^2 + B|\alpha_j^{(k)}| + C - \lambda = 0.$$

Hence $|\alpha_j^{(k)}| = \frac{-B + \sqrt{B^2 - 4A(C-\lambda)}}{2A}$, as $\frac{-B - \sqrt{B^2 - 4A(C-\lambda)}}{2A} < 0$ and is not a valid answer for $|\alpha_j^{(k)}|$. Moreover, as $4A\lambda \geq 0$, to minimize the power $\lambda$ should be equal to zero, i.e, the power should be used to meet the SNR constraint with equality. Hence $|\alpha_j^{(k)}| = \frac{-B + \sqrt{B^2 - 4AC}}{2A}$. Now as $A$ is fixed, to minimize $|\alpha_j^{(k)}|^2$ we should minimize $-B + \sqrt{B^2 - 4AC}$. Note $-B + \sqrt{B^2 - 4AC}$ always has a negative derivative with respect to $B$, hence its minimum is achieved for the maximum value of $B$. Denote $\Gamma_{jk}e^{j\angle\theta_{jk}} = \bar{\mathbf{g}}_{j\mu_{jk}}^H\mathbf{d}_j^{(k)}\mathbf{c}_j^{(k-1)H}\bar{\mathbf{g}}_{j\mu_{jk}}$, we have $B = 2\,\Gamma_{jk}\,\mathrm{Re}(e^{j(\angle\alpha_j^{(k)} + \angle\theta_{jk})})$, the maximum of which achieved if $\angle\alpha_j^{(k)} = -\angle\theta_{jk}$ and $|\alpha_j^{(k)}|$ is given as in (34).